\documentclass[twocolumn,amsmath,amssymb,superscriptaddress,prb,showpacs]{revtex4}

\usepackage{graphicx}

\usepackage[breaklinks]{hyperref}
\hypersetup{
    bookmarks=false,
    pdfstartview={FitH},
    colorlinks=true,
    linkcolor=blue,
    citecolor=blue,
    urlcolor=blue
}

\begin{document}

\title{Photoemission evidence for crossover from Peierls-like to Mott-like
transition in highly strained VO$_2$}

\author{J.~Laverock}
\author{A.~R.~H.~Preston}
\author{D.~Newby, Jr.}
\author{K.~E.~Smith}
\affiliation{Department of Physics, Boston University, 590 Commonwealth Avenue,
Boston, MA 02215, USA}

\author{S.~Sallis}
\author{L.~F.~J.~Piper}
\affiliation{Department of Physics, Applied Physics and Astronomy, Binghamton
University, Binghamton, NY 13902}
%University, Binghamton, NY 13902, USA}

\author{S.~Kittiwatanakul}
\affiliation{Department of Physics, University of Virginia,
Charlottesville, VA 22904, USA}
\author{J.~W.~Lu}
\affiliation{Department of Materials Science and Engineering,
University of Virginia, VA 22904}
%University of Virginia, Charlottesville, VA 22904, USA}
\author{S.~A.~Wolf}
\affiliation{Department of Physics, University of Virginia,
Charlottesville, VA 22904, USA}
\affiliation{Department of Materials Science and Engineering,
University of Virginia, VA 22904}
%University of Virginia, Charlottesville, VA 22904, USA}

\author{M.~Leandersson}
\author{T.~Balasubramanian}
\affiliation{MAX-lab, Lund University, SE-221 00 Lund, Sweden}

% 71.27.+a Strongly correlated electron systems; heavy fermions
% 71.30.+h Metal-insulator transitions and other electronic transitions
% 79.60.-i Photoemission and photoelectron spectra

\begin{abstract}
We present a spectroscopic study that reveals that
the metal-insulator transition of strained VO$_2$ thin films
may be driven towards a purely electronic transition, which does not
rely on the Peierls dimerization, by the application of mechanical strain.
Comparison with a moderately strained
system, which does involve the lattice, demonstrates the crossover
from Peierls- to Mott-like transitions.
%This observation has important
%implications for device physics that rely on the ultrafast nature of VO$_2$,
%and suggests a potential route to circumventing the structural bottleneck.
\end{abstract}

\maketitle

The metal-insulator transition (MIT) in VO$_2$ is of both fundamental and
technical interest, the former due to lingering important questions about its
origins \cite{goodenough1973,zylbersztejn1975},
and the latter due to possible applications in
electronic devices such as ultrafast
optical switches and field effect transistors \cite{rini2005,kim2004}.
In bulk VO$_2$, a large structural
distortion accompanies the transition from the metallic rutile to the
insulating monoclinic phase, which is known to impose a significant
bottleneck on the
timescale of the photo-induced transition \cite{cavalleri2004}.
%We report here
%a spectroscopic study that reveals that
%the MIT may be driven towards a purely electronic transition, which does not
%rely on the Peierls dimerization, by the application of mechanical strain.
%Comparison with a moderately strained
%system, which does involve the lattice, demonstrate the crossover
%from Peierls-like to Mott-like transitions. This observation has important
%implications for device physics that rely on the ultrafast nature of VO$_2$,
%and suggests a potential route to circumventing the structural bottleneck.
%
Recently, the possibility of tailoring the transition temperature of the
MIT in VO$_2$ through doping and/or nanoscale engineering
\cite{muraoka2002,west2008b,cao2009etc}
has heralded renewed interest in the potential
application of VO$_2$ as a novel functional material. Whereas the mechanism
of the MIT in bulk VO$_2$ is now reasonably well understood as an
orbital-assisted collaborative Mott-Peierls transition (or associated
variants) \cite{haverkort2005,weber2012} (i.e.~the MIT involves {\em both}
the lattice and electron-correlation effects), the situation is less clear
under large applied strain to the lattice. For example, the introduction
of Nb as an isoelectronic dopant to V leads to an expansion of the $c/a$
constant, and eventually (for $\geq 15$\% Nb) an insulating rutile phase is
observed \cite{lederer1972} (i.e.~without any structural distortion).

\begin{figure}[b!]
\begin{center}
\includegraphics[width=0.8\linewidth,clip]{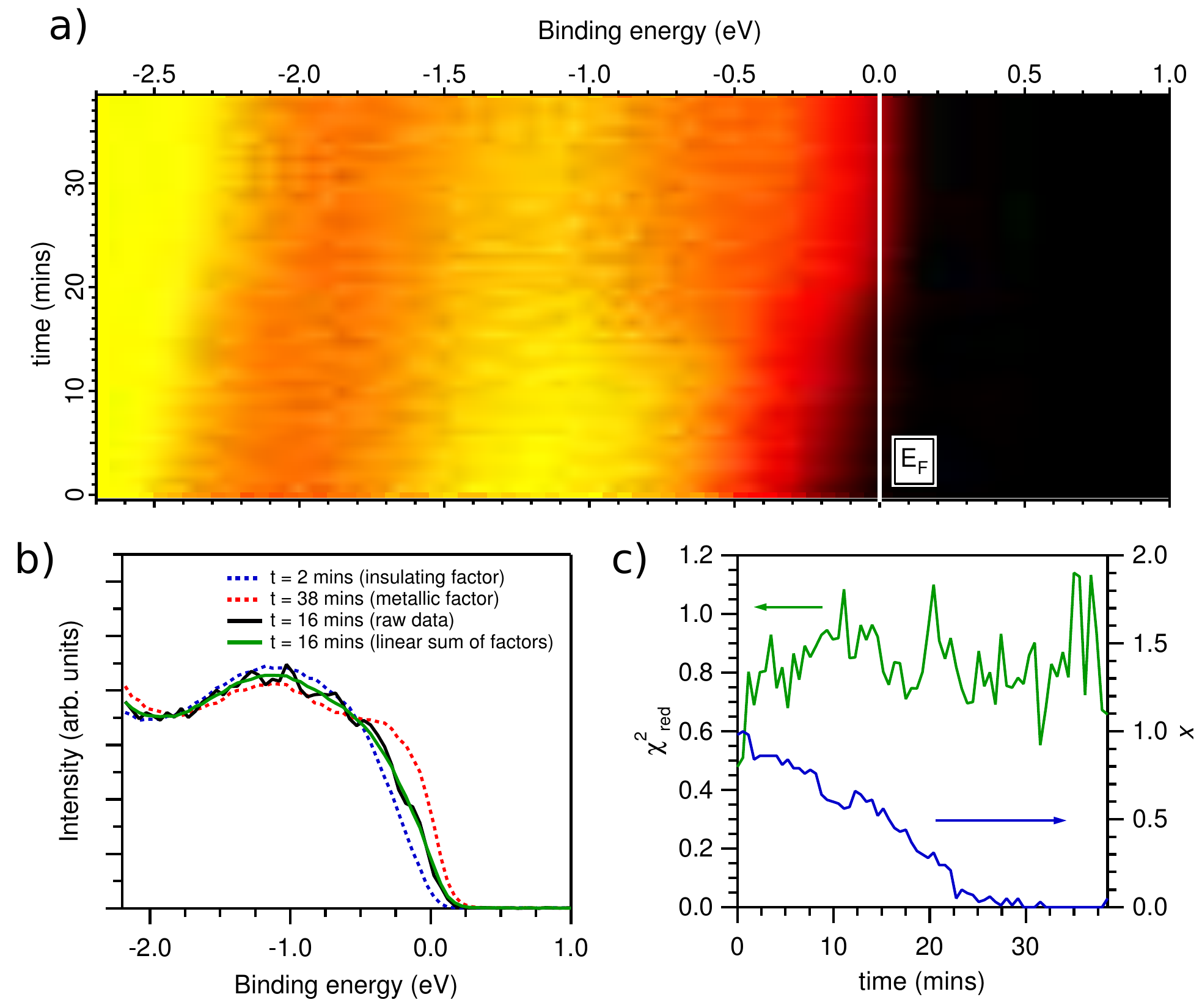}
\end{center}
\vspace*{-0.2in}
\caption{\label{f:heating} V $3d$ PES spectra of
VO$_2$(110) during a heating cycle through the MIT.
(a) A two-dimensional map of the evolution of the V $3d$
states with time.
(b) Spectrum recorded
half-way through the heating cycle compared with a linear sum of the
insulating and metallic factors. (c) The goodness-of-fit parameter,
$\chi^2$, between the time-dependent data and its fit to the two end-members of
the series. Also shown is the evolution of the fraction, $x$, of the insulating
spectrum used to describe the data.}
\end{figure}

The crucial aspects of the electronic structure of VO$_2$ center around
the behavior of the V $a_{1g}$ orbital, which is oriented along the rutile
$c$-axis ($c_{\rm R}$-axis), and is also labeled as $d_{\parallel}$ for this
reason. In the Peierls-type model, this orbital becomes split in the insulating
monoclinic phase due to the dimerization of V atoms in the $c_{\rm R}$-axis,
and the associated twisting of the VO$_6$ octahedra pushes the $e_g^{\pi}$
bands upwards in energy, deoccupying them \cite{goodenough1973}. On
the other hand, the correlation-driven mechanism \cite{zylbersztejn1975}
proposes that the strong correlations in the $a_{1g}$ orbital are screened
by the $e_g^{\pi}$ band in the metallic phase. In the insulating phase,
the $e_g^{\pi}$ states are empty, and the unscreened correlations open
the gap. In the past, it has been difficult to distinguish between these
two models, owing to the large structural distortion that accompanies the
transition. However, a growing body of experimental and theoretical evidence
supports that both are important in driving the MIT of bulk VO$_2$. For
example, the local density approximation (LDA) fails to reproduce the
MIT \cite{eyert2002} without recourse to hybrid functionals beyond the
LDA \cite{eyert2011}, and dynamical mean-field theory (DMFT) calculations
\cite{biermann2005,lazarovits2010} (that explicitly include dynamical electron
correlations) are required to explain a good deal of the experimental data
(for example,
Refs.~\onlinecite{lederer1972,haverkort2005,koethe2006,muraoka2002}).
%However, accessing the {\bf k}-resolved electron dispersion relations
%of VO$_2$ with
%photoemission has proven problematic in the past, in particular owing to the
%difficulty in preparing atomically clean, well-ordered and chemically stable
%surfaces.
We report here a photoemission spectroscopy (PES) study of strained VO$_2$
thin films, in which we show that the MIT may be driven towards a purely
electronic transition, which does not rely on the Peierls dimerization, by the
application of mechanical strain.
%where we take advantage
%of the polarisability of the incident synchrotron radiation photons,
%coupled with the orbital
%anisotropy of the $t_{2g}$ orbitals, to probe the electronic structure
%of two representative systems.

\begin{figure}[t!]
\begin{center}
\includegraphics[width=0.9\linewidth,clip]{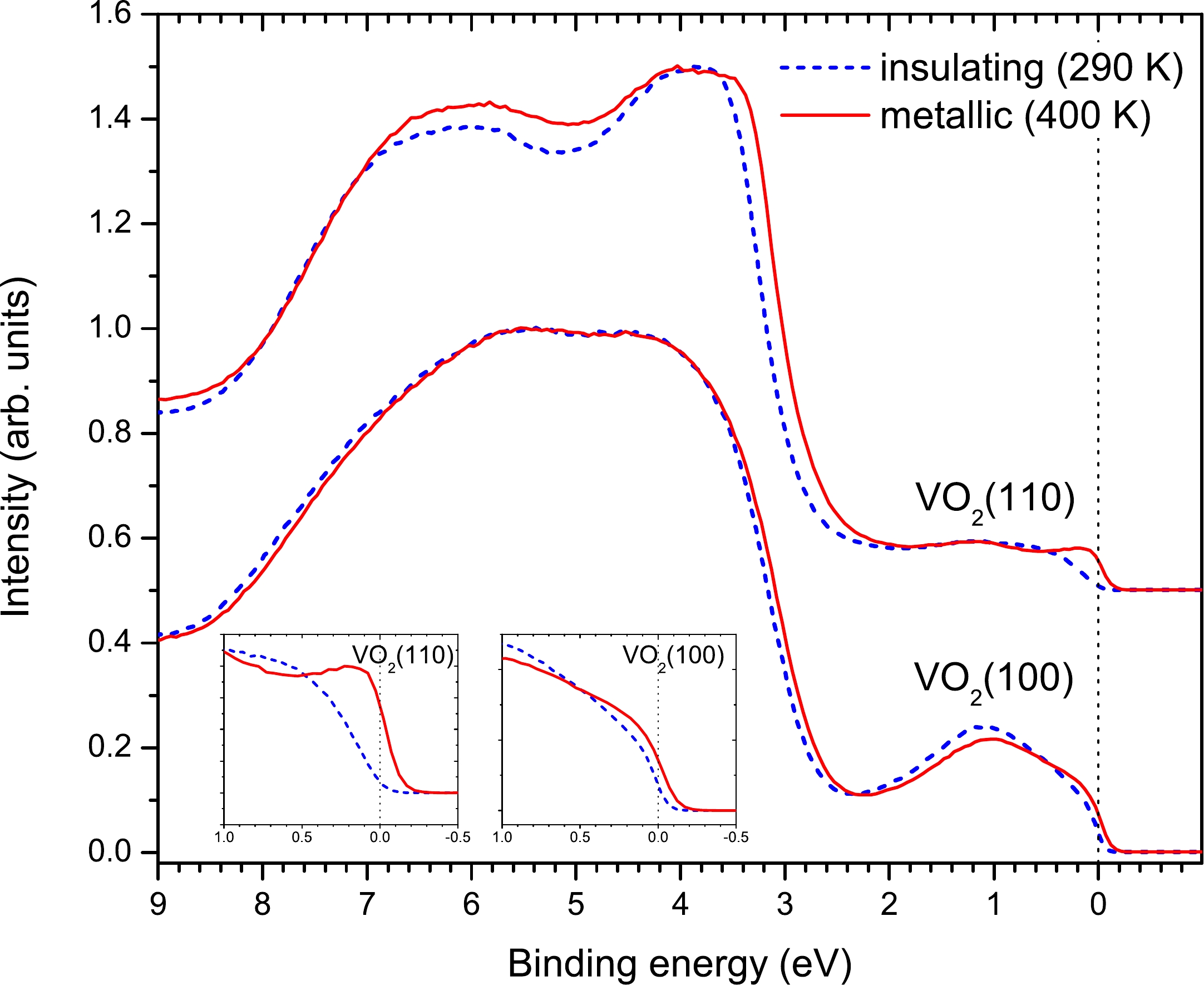}
\end{center}
\vspace*{-0.2in}
\caption{\label{f:inout} PES spectra of the O $2p$ and V $3d$
states across the MIT of strained VO$_2$, recorded at 30~eV and with $E
\parallel c_{\rm R}$. The insets
show an enlargement near $E_{\rm F}$.}
\end{figure}

High-quality thin films ($\sim 40$~nm) of VO$_2$ were
grown on rutile TiO$_2$(110)- and TiO$_2$(100)-oriented substrates by
reactive bias target ion-beam deposition \cite{west2008b}. X-ray diffraction
measurements confirm the epitaxy of VO$_2$ with the substrate and establish
the expanded $c_{\rm R}$-axis lattice parameter of VO$_2$ compared with
the bulk \cite{suppl}.
This tensive strain is found to be approximately twice as large
for VO$_2$/TiO$_2$(100) [hereafter referred to as VO$_2$(100)]
at $+3.7\%$, whereas the $c_{\rm R}$-axis strain
of VO$_2$/TiO$_2$(110) [referred to as VO$_2$(110)] is $+1.7\%$.
Atomic force microscopy
measurements \cite{suppl} suggest a very low root-mean-square roughness of
$\sim 0.3$~--~0.4~nm.
Both samples display MITs above room temperature, at 345~K and 340~K
for VO$_2$(110) and VO$_2$(100) respectively. Note that the MIT temperature of
our VO$_2$(110) film is lower than that observed in
Ref.~\onlinecite{muraoka2002}.
PES measurements were carried out at beamline I3 (MAX III) of
MAX-lab, Lund University, Sweden, using a Scienta R-4000 analyzer set to an
energy resolution of 12~meV, with pressures in the analysis chamber of better
than $5 \times 10^{-10}$~torr between room temperature and $150^{\circ}$C. The
binding energy of the PES spectra were referenced
to polycrystalline gold in electrical contact with the samples. The samples
were prepared for ultra-high vacuum measurements by several (one to three)
repetitions of annealing in a partial O$_2$ environment ($\sim 450^{\circ}$C,
$1 \times 10^{-6}$~torr O$_2$, 30~mins). This procedure has been carefully
developed to remove contaminant surface species from the samples without
modifying the chemical environment of VO$_2$. Samples were aligned using
low-energy electron diffraction and Laue x-ray diffraction. Soft x-ray
spectroscopy measurements were performed at beamline X1B of the National
Synchrotron Light Source (NSLS), Brookhaven.  X-ray absorption
spectroscopy (XAS) measurements
were made in total electron yield mode with a beamline resolution of
0.2~eV, and the photon energy was calibrated with reference to TiO$_2$ spectra.
Reonsant x-ray emission spectroscopy (RXES) measurements were recorded with
a Nordgren-type grating spectrometer \cite{nordgren1989}
set to an energy resolution of 0.7~eV and the
instrument was calibrated using a Zn reference spectrum.

For {\em moderately-strained} VO$_2$(110),
angle-integrated PES spectra of the V $3d$ states near $E_{\rm
F}$ are shown in Fig.~\ref{f:heating}a as a function of time during
a slow heating cycle through the MIT. The evolution of the V $3d$ states
is characterized by a transfer in spectral weight from $\sim 1$~eV binding
energy towards $E_{\rm F}$, associated with the shift in the quasiparticle
peak energy that becomes gapped out below the MIT \cite{biermann2005}. The
small remnant weight near 1~eV in the metallic phase represents the lower
Hubbard band (LHB) of the rutile metallic phase, and is slightly shifted towards
higher binding energies, in agreement with cluster DMFT (cDMFT)
results \cite{biermann2005}. These results are in
excellent agreement with previous PES measurements of both bulk and
thin-film VO$_2$ \cite{koethe2006,okazaki2004etc,eguchi2008etc}.
In Fig.~\ref{f:heating}b, a comparison of the metallic and insulating
spectra (end members of the heating cycle) is made with a spectrum recorded
mid-way through the series. A linear superposition of the two end members
is also shown, supporting
real space measurements of bulk VO$_2$ that find no evidence
of an intermediate phase in the transition \cite{corr2010}. Indeed,
the spectra throughout the heating cycle have been carefully analyzed using
a factor analysis approach, and we find only two factors are required to
reproduce the spectra for all energies and temperatures measured
(Fig.~\ref{f:heating}c).

\begin{figure*}[t!]
\begin{center}
\includegraphics[width=1.\linewidth,clip]{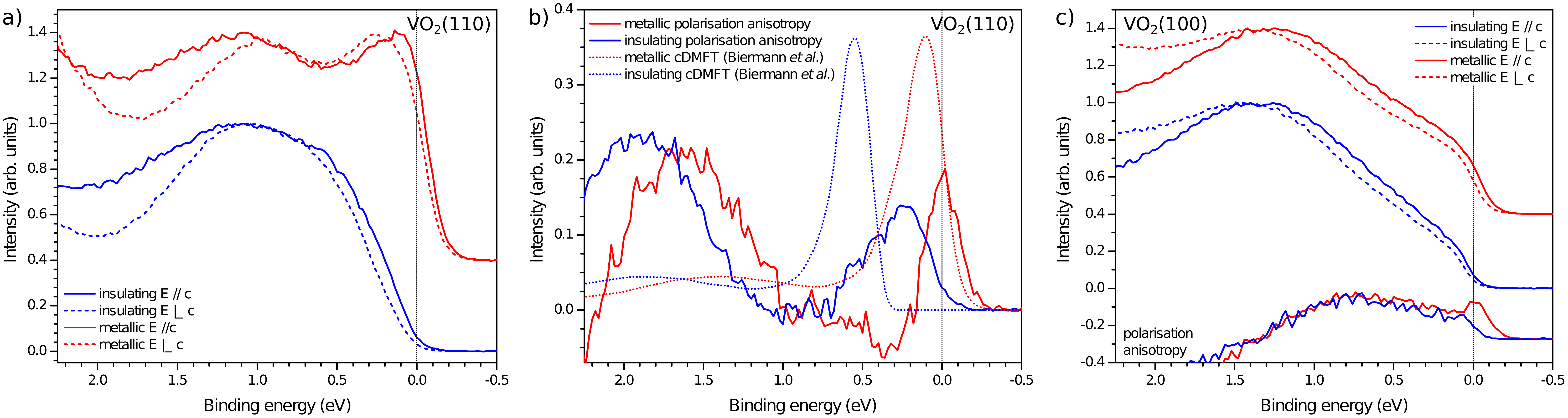}
\end{center}
\vspace*{-0.2in}
\caption{\label{f:pani} High resolution V $3d$ PES spectra across
the MIT of the VO$_2$ thin films at 30~eV. (a) Spectra of moderately-strained
VO$_2$(110) for two different incident photon polarizations: ${E} \parallel
c_{\rm R}$ and ${E} \perp c_{\rm R}$. (b) Polarization anisotropy of
VO$_2$(110) in the insulating and metallic phases compared with cDMFT
calculations of Biermann {\em et al.}~\cite{biermann2005}.  (c)
Polarization-dependent spectra of
highly-strained VO$_2$(100).
Shown at the bottom of the figure is the weak polarization anisotropy for
this system.}
\end{figure*}

In Fig.~\ref{f:inout}, angle-integrated PES spectra
of the O $2p$ and V $3d$ states
of moderately strained VO$_2$(110) and VO$_2$(100) are shown across the MIT.
In VO$_2$(110), a large gap develops
in the insulating (room temperature) spectra, the magnitude of which is
$\sim 300$~meV (measured from extrema in the first derivative). However,
for the {\em highly strained} VO$_2$(100) system (Fig.~\ref{f:inout}b),
the magnitude of the insulating gap is much smaller, at $\lesssim 50$~meV,
with only a weak shift in the O $2p$ band.  Here, the strong double-peaked
structure in the O $2p$ manifold is less prominent; rather the spectra exhibit
broad, asymmetric peaks, similar to earlier PES studies of VO$_2$
\cite{okazaki2004etc}. The weak
opening of the insulating gap in films under high strain compared to moderate
strain
constitutes our first clue that the physics of the MIT may be different
for highly strained VO$_2$(100).

In order to understand in more detail the behavior of the two systems
across the MIT, we now focus on the dependence of the V $3d$ states with
photon polarization.
By rotating the polarization vector of the
incident photons, it is possible to couple to different symmetry orbitals,
providing orbital-resolution to PES measurements. In particular,
for ${E} \parallel c_{\rm R}$, the matrix elements that
couple the PES process of the V $3d$ states are maximized for the
$a_{1g}$ states.
In Fig.~\ref{f:pani}a, high-resolution
angle-integrated spectra above
and below the MIT are shown for ${E} \parallel c_{\rm R}$ and ${E}
\perp c_{\rm R}$ for VO$_2$(110), normalized to the intensity between 1.0 and
1.2~eV. It is clear from Fig.~\ref{f:pani}a that there
is a substantial change in the shape of the spectra across the transition, and
even in the relative energies of features between different polarizations. In
Fig.~\ref{f:pani}b, the polarization anisotropy, $I_{\parallel} -
I_{\perp}$, is shown for the spectra from metallic and insulating phases,
representing
approximate experimental PES spectra of the $a_{1g}$
orbital. In the metallic phase, the quasiparticle peak is centered at $E_{\rm
F}$, with the LHB at $\sim 1.5$~eV.  The quasiparticle peak
shifts down to $\sim 0.4$~eV in the insulating spectra,
accompanied by a shift of the LHB to $\sim 1.8$~eV. These results are in
remarkable agreement with the cDMFT calculations of the $a_{1g}$
orbital of Biermann {\em et al.}~\cite{biermann2005}, which are reproduced
by the dotted lines in Fig.~\ref{f:pani}b. In the $M_1$
insulating phase of VO$_2$, the $a_{1g}$ state is overwhelmingly the
most occupied of the V $t_{2g}$ orbitals (contributing $\sim 80$\% to the total
occupied density of states), whereas these orbitals are almost
evenly populated in the metallic rutile phase.
Further evidence of this interpretation is provided
by parameter-free analysis of the insulating spectra.
For ${E} \perp c_{\rm R}$, the extremum of the first derivative
is located $\sim 300$~meV below $E_{\rm F}$, but is much smaller for ${E}
\parallel c_{\rm R}$ at $\sim 150$~meV (the associated error in determining
this quantity is $\sim 10$~meV), supporting the cDMFT results that show the
$a_{1g}$ orbital lies closer to $E_{\rm F}$ than the $e_g^{\pi}$ states.

\begin{figure}[b!]
\begin{center}
\includegraphics[width=0.7\linewidth,clip]{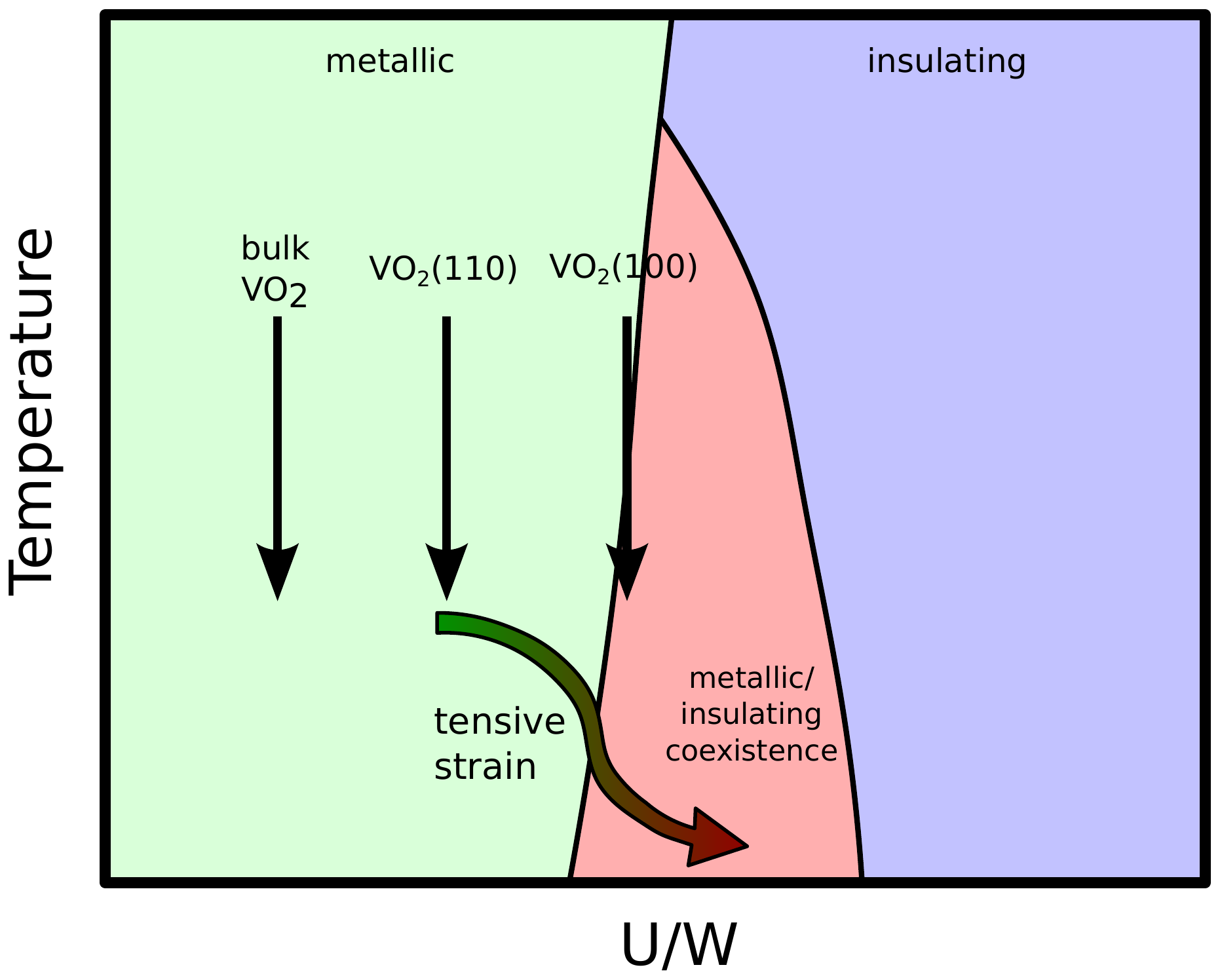}
\end{center}
\vspace*{-0.2in}
\caption{\label{f:cartoon} Schematic representation of the impact of tensive
$c_{\rm R}$-axis strain on the phase diagram of VO$_2$. Tensive strain expands
the $c_{\rm R}$-axis, narrowing the bandwidth, $W$, and leading to an increase
in the relative importance of electron correlations, $U/W$. The phase diagram is
based on the DMFT results of Ref.~\onlinecite{lazarovits2010}.}
\end{figure}

\begin{figure*}[t!]
\begin{center}
\includegraphics[width=0.7\linewidth,clip]{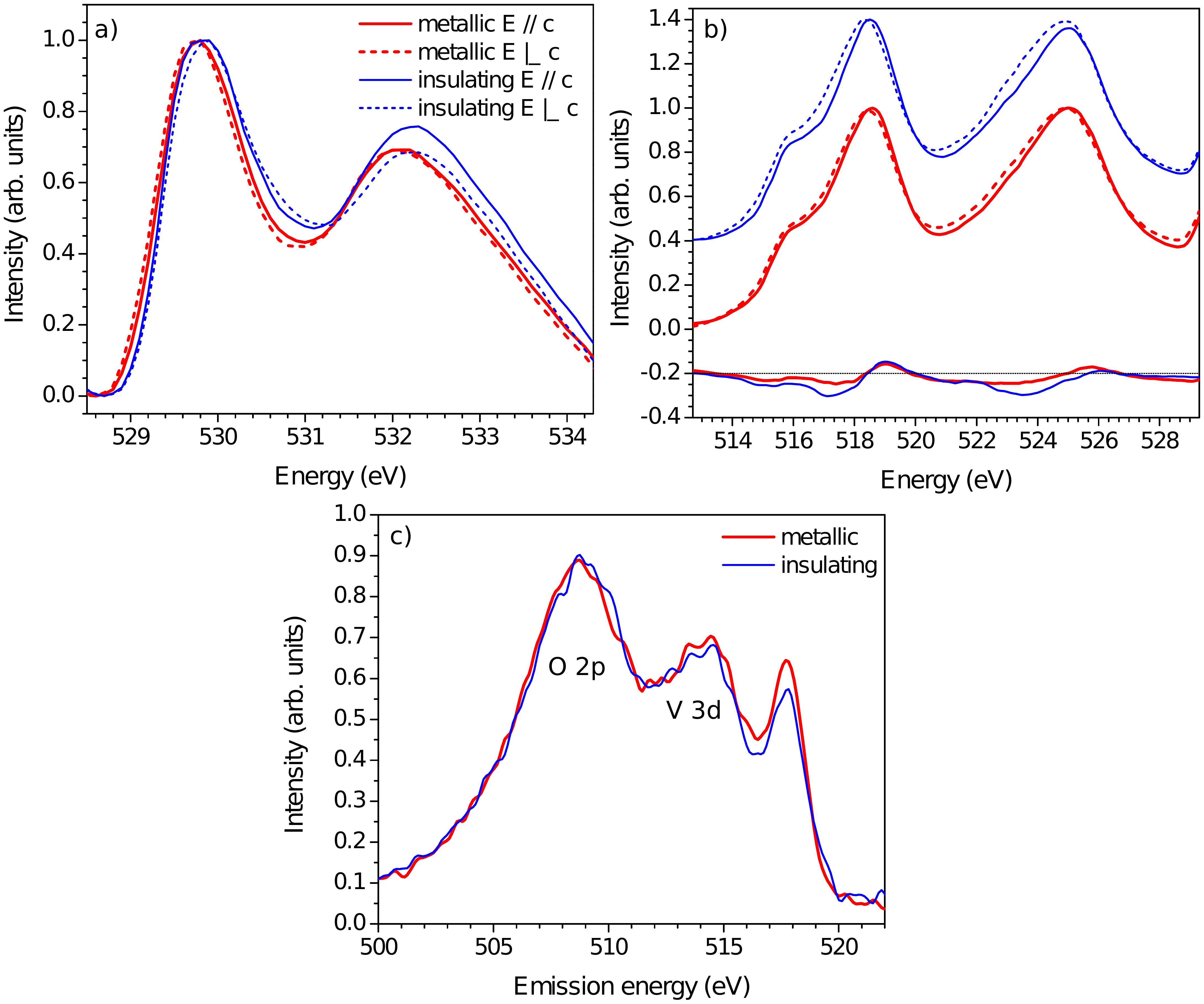}
\end{center}
\vspace*{-0.2in}
\caption{\label{f:xxs} Soft x-ray measurements on VO$_2$(100) through the MIT.
(a) Polarization-dependent O $K$-edge XAS.
(b) Polarization-dependent V $L_{3,2}$-edge XAS.
At the bottom of the figure, the anisotropy between ${E}
\parallel c_{\rm R}$ and ${E} \perp c_{\rm R}$ are shown. (c) RXES
spectra, recorded at the V $L_3$-edge peak.}
\end{figure*}

In Fig.~\ref{f:pani}c, high-resolution spectra of highly-strained VO$_2$(100)
are shown alongside their polarization anisotropy. These spectra show
much weaker anisotropy, with very similar shapes both above and below the
transition: a very
small quasiparticle `peak' shifts to deeper energies (and diminishes in
intensity) in the insulating phase. Examination of the first derivative of
both photon polarizations in the insulating phase reveal both extrema lie
$\sim 50$~meV below $E_{\rm F}$, demonstrating the relative isotropy of the
gap for VO$_2$(100). The similarity of the polarization anisotropy across the
MIT suggests that the population of orbitals is approximately the same in the
insulating and metallic phases, inconsistent with the structural distortion
that preferentially occupies the $a_{1g}$ orbitals. DMFT calculations of the
rutile phase with varying on-site Hubbard $U$ have demonstrated that an {\em
insulating rutile} phase (i.e.~without structural distortion) may be stabilized
for large values of $U$ \cite{lazarovits2010}. Bulk rutile VO$_2$ is believed
to lie close to (but on the metallic side of) the crossover region (in which
both metallic and insulating phases are stable). However, the nature of the
insulating rutile phase is very different from the insulating $M_1$ phase of
bulk VO$_2$, and is characterized by a very small insulating gap and almost
even population of $t_{2g}$ orbitals (similar to the metallic phase). Our
PES measurements of highly strained VO$_2$(100) are consistent
with such a scenario, in which we observe both a very small insulating gap
($\lesssim 50$~meV) {\em and} similar populations of the $t_{2g}$ orbitals
across the transition, in contrast to our results of the moderately strained
VO$_2$(110) system. It is suggested that the
large tensive strain of VO$_2$(100) leads to a narrowing of the bandwidth, $W$,
of the $t_{2g}$ bands, and the accompanying increase in the relative importance
of electron correlations, characterized by $U/W$. In this simple model
(schematically depicted in Fig.~\ref{f:cartoon}),
$U/W_{\rm bulk} < U/W_{(110)} < U/W_{(100)}$, and the system may be pushed
into the coexistence region of the phase diagram, with a crossover from the
traditional lattice-electronic mechanism of the bulk system to an electronic
one for the highly strained system. For Nb-doped VO$_2$, an insulating rutile
phase has already been observed, for which the mechanism was interpreted as
electronic in origin, assisted by the disorder \cite{lederer1972}. Here,
we show that such an electronic driven transition can also be stabilized
for pure VO$_2$ at ambient pressures by introducing large tensive strain
to the system. These findings are in agreement with a recent optical study
of similar VO$_2$ thin films, in which the decoupling of the structural and
electronic components of the MIT was reported \cite{abreu2012}.  One can
then circumvent the well-known structural bottleneck in the timescale of
the transition \cite{cavalleri2004}, of substantial value for applications
that are envisaged to exploit the ultra-fast nature of the MIT of VO$_2$
\cite{rini2005,cavalleri2001}. Furthermore, this observation that the
electronic-driven transition in highly-strained VO$_2$ is quite different
to that in moderately strained VO$_2$ (in agreement with DMFT) supplies
additional weight to the argument that {\em both} the Peierls-type and
Mott-type mechnisms are important in stabilizing the $M_1$ phase of bulk
VO$_2$.

In order to reinforce our PES results, soft x-ray XAS and
RXES measurements have also been performed, and are shown in
Fig.~\ref{f:xxs}. It
is well known that XAS of the O $K$-edge
\cite{abbate1991,koethe2006} and V $L_{3,2}$-edge \cite{haverkort2005}
are sensitive to changes in the population of the $t_{2g}$ orbitals
across the MIT, and we have previously demonstrated that RXES
at the V $L_3$-edge is a sensitive probe of
the structural distortion associated with the MIT in moderately strained
VO$_2$ \cite{laverock2012,piper2010b}. For VO$_2$(100), we find
our XAS spectra from films in the metallic phase at the V $L_{3,2}$-
and O $K$-edges (Fig.~\ref{f:xxs}a,b) are in good agreement with other such
measurements of the metallic phase of bulk and moderately-strained VO$_2$
\cite{haverkort2005,koethe2006,laverock2012}. However, the spectra from
insulating films are almost identical to those
from the metallic films, with only a weak shift in
the O $K$-edge threshold (of $\lesssim 0.1$~eV, which is the accuracy of
these measurements) indicating the very small insulating gap. In particular,
the $a_{1g}$ ($d_{\parallel}$) peak, a signature of V-V dimerization
in the distorted monoclinic phase, is absent at low temperature.
Similarly, RXES measurements of VO$_2$(100) are not found to exhibit any
temperature dependence across the MIT (Fig.~\ref{f:xxs}c).
For moderately strained VO$_2$(110), a large change in the
relative intensity of the V $3d$ feature at $\sim 515$~eV is related to the
different bonding environments of V-O induced by the structural distortion
of the VO$_6$ octahedra
at the MIT \cite{laverock2012}. Here, we find this ratio is the same for the
insulating and metallic spectra of VO$_2$(100), indicating the twisting of the
VO$_6$ octahedra is absent in this sample.
Both of these observations
are consistent with the PES evidence described above, and together
suggest a
rutile-like structure exists in both metallic and insulating phases.
Moreover, recent temperature-dependent x-ray diffraction measurements on
similar thin films \cite{abreu2012} suggest that the change in lattice
spacing at the MIT is very weak (by an order of magnitude) for their highly
strained
VO$_2$(100) films compared with the bulk, and is decoupled from the electronic
transition. For our VO$_2$(100) samples, we find a similar suppression of the
lattice spacing component of the transition compared with VO$_2$(110)
\cite{suppl}, emphasizing the weaker role of the lattice in the MIT of
VO$_2$(100).

In summary, we have observed a crossover from a Mott-Peierls-like transition
to a Mott-like transition with an increase in tensive strain along the $c_{\rm
R}$-axis in VO$_2$ through several different spectroscopic techniques. XAS
and RXES were used to demonstrate the
absence of the large structural distortion at the MIT that characterizes bulk
and moderately strained VO$_2$. PES measurements revealed
a weak insulating gap as well as the suppression of orbital redistribution
across the transition. We further showed that by exploiting the relative
polarization of the
incident photons, PES can probe changes in the
orbital occupation across electronic phase transitions.  Our observations
have important implications for novel functional material engineering of
VO$_2$, suggesting a route towards circumventing the structural bottleneck
in the ultrafast timescale of the MIT.

{\bf Acknowledgements}. The authors acknowledge useful discussions with
R.~D.~Averitt, E.~Abreu
and M.~Liu. The Boston University program is supported in part
by the Department of Energy under Grant No.\ DE-FG02-98ER45680. The NSLS,
Brookhaven, is supported by the U.S.\ Department of Energy under Contract
No.\ DE-AC02-98CH10886. SK, JWL and SAW are thankful for financial support
from the Army Research Office through MURI Grant No.\ W911-NF-09-1-0398.


\begin{thebibliography}{99}

\bibitem{goodenough1973}
J.\ B.\ Goodenough and H.\ Y.-P.\ Hong,
\href{http://dx.doi.org/10.1103/PhysRevB.8.1323}
{Phys.\ Rev.\ B {\bf 8}, 1323 (1973)}.

\bibitem{zylbersztejn1975}
A.\ Zylbersztejn and N.\ F.\ Mott,
\href{http://dx.doi.org/10.1103/PhysRevB.11.4383}
{Phys.\ Rev.\ B {\bf 11} 4383 (1975)}.

\bibitem{rini2005}
%M.\ Rini {\it et al.},
M.\ Rini, A.\ Cavalleri, R.\ W.\ Schoenlein, R.\ L\'{o}pez, L.\ C.\ Feldman,
R.\ F.\ Haglund, Jr., L.\ A.\ Boatner and T.\ E.\ Haynes,
\href{http://dx.doi.org/10.1364/OL.30.000558}
{Optics Lett.\ {\bf 30}, 558 (2005)}.

\bibitem{kim2004}
%H.-T.\ Kim {\it et al.},
H.-T.\ Kim, B.-G.\ Chae, D.-H.\ Youn, S.-L.\ Maeng, G.\ Kim, K.-Y.\ Kang
and Y.-S.\ Lim,
\href{http://dx.doi.org/10.1088/1367-2630/6/1/052}
{New J.\ Phys.\ {\bf 6}, 52 (2004)}.

\bibitem{cavalleri2004}
%A.\ Cavalleri {\it et al.},
A.\ Cavalleri, Th.\ Dekorsy, H.\ H.\ W.\ Chong, J.\ C.\ Kieffer and R.\ W.\
Schoenlein,
\href{http://dx.doi.org/10.1103/PhysRevB.70.161102}
{Phys.\ Rev.\ B {\bf 70}, 161102(R) (2004)}.

\bibitem{muraoka2002}
Y.\ Muraoka and Z.\ Hiroi,
\href{http://dx.doi.org/10.1063/1.1446215}
{Appl.\ Phys.\ Lett.\ {\bf 80}, 583 (2002)}.

\bibitem{west2008b}
%K.\ G.\ West {\it et al.},
K.\ G.\ West, J.\ W.\ Lu, J.\ Yu, D.\ Kirkwood, W.\ Chen, Y.\ H.\ Pei,
J.\ Claassen and S.\ A.\ Wolf,
\href{http://dx.doi.org/10.1116/1.2819268}
{J.\ Vac.\ Sci.\ Technol.\ A {\bf 26}, 133 (2008)}.

\bibitem{suppl}
See Supplemental Material at [url] for details of the sample growth and
characterization.

\bibitem{cao2009etc}
%J.\ Cao {\it et al.},
J.\ Cao, E.\ Ertekin, V.\ Srinivasan, W.\ Fan, S.\ Huang, H.\ Zheng,
J.\ W.\ L. Yim, D.\ R.\ Khanal, D.\ F.\ Ogletree, J.\ C.\ Grossman and J.\ Wu,
\href{http://dx.doi.org/10.1038/nnano.2009.266}
{Nature Nanotech.\ {\bf 4}, 732 (2009)};
%
%\bibitem{cao2010}
%J.\ Cao {\it et al.},
J.\ Cao, Y.\ Gu, W.\ Fan, L.\ Q.\ Chen, D.\ F.\ Ogletree, K.\ Chen, N.\ Tamura,
M.\ Kunz, C.\ Barrett, J.\ Seidel and J.\ Wu,
\href{http://dx.doi.org/10.1021/nl101456k}
{Nano Lett.\ {\bf 10}, 2667 (2010)}.

\bibitem{haverkort2005}
%M.\ W.\ Haverkort {\it et al.},
M.\ W.\ Haverkort, Z.\ Hu, A.\ Tanaka, W.\ Reichelt, S.\ V.\ Streltsov, M.\ A.\
Korotin, V.\ I.\ Anisimov, H.\ H.\ Hsieh, H.-J.\ Lin, C.\ T.\ Chen, D.\ I.\
Khomskii and L.\ H.\ Tjeng,
\href{http://dx.doi.org/10.1103/PhysRevLett.95.196404}
{Phys.\ Rev.\ Lett.\ {\bf 95}, 196404 (2005)}.

\bibitem{weber2012}
%C.\ Weber {\it et al.},
C.\ Weber, D.\ D.\ O'Regan, N.\ D.\ M.\ Hine, M.\ C.\ Payne, G.\ Kotliar and
P.\ B.\ Littlewood,
\href{http://dx.doi.org/10.1103/PhysRevLett.108.256402}
{Phys.\ Rev.\ Lett.\ {\bf 108}, 256402 (2012)}.

\bibitem{lederer1972}
%P.\ Lederer {\it et al.},
P.\ Lederer, H.\ Launois, J.\ P.\ Pouget, A.\ Casalot and G.\ Villeneuve,
\href{http://dx.doi.org/10.1016/S0022-3697(72)80496-7}
{J.\ Phys.\ Chem.\ Solids {\bf 33}, 1969 (1972)}.

\bibitem{eyert2002}
V.\ Eyert,
\href{http://dx.doi.org/10.1002/1521-3889(200210)11:9<650::AID-ANDP650>3.0.CO;2-K}
{Ann.\ Phys.\ (Leipzig) {\bf 11}, 650 (2002)}.

\bibitem{eyert2011}
V.\ Eyert,
\href{http://dx.doi.org/10.1103/PhysRevLett.107.016401}
{Phys.\ Rev.\ Lett.\ {\bf 107}, 016401 (2011)}.

\bibitem{biermann2005}
%S.\ Biermann {\it et al.},
S.\ Biermann, A.\ Poteryaev, A.\ I.\ Lichtenstein and A.\ Georges,
\href{http://dx.doi.org/10.1103/PhysRevLett.94.026404}
{Phys.\ Rev.\ Lett.\ {\bf 94}, 026404 (2005)}.

\bibitem{lazarovits2010}
%B.\ Lazarovits {\it et al.},
B.\ Lazarovits, K.\ Kim, K.\ Haule and G.\ Kotliar,
\href{http://dx.doi.org/10.1103/PhysRevB.81.115117}
{Phys.\ Rev.\ B {\bf 81}, 115117 (2010)}.

\bibitem{koethe2006}
%T.\ C.\ Koethe {\it et al.},
T.\ C.\ Koethe, Z.\ Hu, M.\ W.\ Haverkort, C.\ Sch\"{u}{\ss}ler-Langeheine, F.\
Venturini, N.\ B.\ Brookes, O.\ Tjernberg, W.\ Reichelt, H.\ H.\ Hsieh, H.-J.\
Lin, C.\ T.\ Chen and L.\ H.\ Tjeng,
\href{http://dx.doi.org/10.1103/PhysRevLett.97.116402}
{Phys.\ Rev.\ Lett.\ {\bf 97}, 116402 (2006)}.

\bibitem{nordgren1989}
J.\ Nordgren, G.\ Bray, S.\ Cramm, R.\ Nyholm, J.-E.\ Rubensson and N.\
Wassdahl,
\href{http://dx.doi.org/10.1063/1.1140929}
{Rev.\ Sci.\ Instrum.\ {\bf 60}, 1690 (1989)}.

\bibitem{okazaki2004etc}
%K.\ Okazaki {\it et al.},
K.\ Okazaki, H.\ Wadati, A.\ Fujimori, M.\ Onoda, Y.\ Muraoka and Z.\ Hiroi,
\href{http://dx.doi.org/10.1103/PhysRevB.69.165104}
{Phys.\ Rev.\ B {\bf 69}, 165104 (2004)}
%\bibitem{saeki2009}
%K.\ Saeki {\it et al.},
K.\ Saeki, T.\ Wakita, Y.\ Muraoka, M.\ Hirai, T.\ Yokoya, R.\ Eguchi and
S.\ Shin,
\href{http://dx.doi.org/10.1103/PhysRevB.80.125406}
{Phys.\ Rev.\ B {\bf 80}, 125406 (2009)}.

\bibitem{eguchi2008etc}
%R.\ Eguchi {\it et al.},
R.\ Eguchi, M.\ Taguchi, M.\ Matsunami, K.\ Horiba, K.\ Yamamoto, Y.\ Ishida,
A.\ Chainani, Y.\ Takata, M.\ Yabashi, D.\ Miwa, Y.\ Nishino, K.\ Tamasaku,
T.\ Ishikawa, Y.\ Senba, H.\ Ohashi, Y.\ Muraoka, Z.\ Hiroi and S.\ Shin,
\href{http://dx.doi.org/10.1103/PhysRevB.78.075115}
{Phys.\ Rev.\ B {\bf 78}, 075115 (2008)};
%\bibitem{kanki2011}
%T.\ Kanki {\it et al.},
T.\ Kanki, H.\ Takami, S.\ Ueda, A.\ N.\ Hattori, K.\ Hattori, H.\ Daimon,
K.\ Kobayashi and H.\ Tanaka,
\href{http://dx.doi.org/10.1103/PhysRevB.84.085107}
{Phys.\ Rev.\ B {\bf 84}, 085107 (2011)}.

\bibitem{corr2010}
%S.\ A.\ Corr {\it et al.},
S.\ A.\ Corr, D.\ P.\ Shoemaker, B.\ C.\ Melot and R.\ Seshadri,
\href{http://dx.doi.org/10.1103/PhysRevLett.105.056404}
{Phys.\ Rev.\ Lett.\ {\bf 105}, 056404 (2010)}.

\bibitem{abreu2012}
%E.\ Abreu {\it et al.},
E.\ Abreu, M.\ Liu, J.\ Lu, K.\ G.\ West, S.\ Kittiwatanakul, W.\ Yin,
S.\ A.\ Wolf and R.\ D.\ Averitt,
\href{http://dx.doi.org/10.1088/1367-2630/14/8/083026}
{New J.\ Phys.\ {\bf 14}, 083026 (2012)}.

\bibitem{cavalleri2001}
%A.\ Cavalleri {\it et al.},
A.\ Cavalleri, Cs.\ T\'{o}th, C.\ W.\ Siders, J.\ A.\ Squier, F.\ R\'{a}ksi,
P.\ Forget and J.\ C.\ Kieffer,
\href{http://dx.doi.org/10.1103/PhysRevLett.87.237401}
{Phys.\ Rev.\ Lett.\ {\bf 87}, 237401 (2001)}.

\bibitem{abbate1991}
%M.\ Abbate {\it et al.},
M.\ Abbate, F.\ M.\ F.\ de Groot, J.\ C.\ Fuggle, Y.\ J.\ Ma, C.\ T.\ Chen, F.\
Sette, A.\ Fujimori, Y.\ Ueda and K.\ Kosuge,
\href{http://dx.doi.org/10.1103/PhysRevB.43.7263}
{Phys.\ Rev.\ B {\bf 43}, 7263 (1991)}.

\bibitem{laverock2012}
%J.\ Laverock {\it et al.},
J.\ Laverock, L.\ F.\ J.\ Piper, A.\ R.\ H.\ Preston, B.\ Chen, J.\ McNulty,
K.\ E.\ Smith, S.\ Kittiwatanakul, J.\ W.\ Lu, S.\ A.\ Wolf, P.-A.\ Glans and
J.-H.\ Guo,
\href{http://dx.doi.org/10.1103/PhysRevB.85.081104}
{Phys.\ Rev.\ B {\bf 85}, 081104(R) (2012)}.

\bibitem{piper2010b}
%L.\ F.\ J.\ Piper {\it et al.},
L.\ F.\ J.\ Piper, A.\ DeMasi, S.\ W.\ Cho, A.\ R.\ H.\ Preston, J.\ Laverock,
K.\ E.\ Smith, K.\ G.\ West, J.\ W.\ Lu and S.\ A.\ Wolf,
\href{http://dx.doi.org/10.1103/PhysRevB.82.235103}
{Phys.\ Rev.\ B {\bf 82}, 235103 (2010)}.

\end{thebibliography}
\end{document}